\documentclass{aa}
\usepackage{graphicx}
\usepackage{txfonts}
\usepackage{natbib}
\begin{document}
\title{New high-resolution near-infrared observations of the asymmetric jet of the massive young stellar object G192.16-3.82}
\titlerunning{High-resolution near-IR observations of G192.16-3.82}
\authorrunning{P. A. Boley et al.}
\author{Paul A. Boley
    \inst{1,2,3}
    \and
    Hendrik Linz
    \inst{2}
    \and
    Nadezhda Dmitrienko
    \inst{1}
    \and
    Iskren Y. Georgiev
    \inst{2}
    \and
    Sebastian Rabien
    \inst{4}
    \and
    Lorenzo Busoni
    \inst{5}
    \and
    Wolfgang G\"assler
    \inst{2}
    \and
    Marco Bonaglia
    \inst{5}
    \and
    Gilles Orban de Xivry
    \inst{6,4}
    }
\institute{Moscow Institute of Physics and Technology, 9 Institutskiy per., 141701 Dolgoprudny, Moscow Region, Russia
    \newline
    \email{boley.pa@mipt.ru,dmitrienko.ns@phystech.edu}
    \and
    Max Planck Institute for Astronomy, K\"onigstuhl 17, 69117 Heidelberg, Germany
    \newline
    \email{\{linz,georgiev,gaessler\}@mpia.de}
    \and
    Ural Federal University, 19 Mira Str., 620075 Ekaterinburg, Russia
    \and
    Max Planck Institute for Extraterrestrial Physics, Gießenbachstraße 1, 85748 Garching, Germany
    \newline
    \email{srabien@mpe.mpg.de}
    \and
    INAF - Arcetri Astrophysical Observatory, Largo Enrico Fermi 5, I-50125 Florence, Italy
    \newline
    \email{\{lbusoni,bonaglia\}@arcetri.inaf.it}
    \and
    Space sciences, Technologies, and Astrophysics Research (STAR) Institute, University of Li\`ege, Place du 20-Août, 7B-4000 Li\`ege, Belgium
    \newline
    \email{gorban@uliege.be}
    }

\date{Draft from \today}

\abstract{The process of massive star formation is tightly connected with the appearance of molecular outflows, which interact with surrounding interstellar medium and can be used as a proxy to study the accretion process of material onto forming massive stars.}{We aim to characterize the morphology and kinematics, as well as the driving source, of the molecular outflow from the massive young stellar object G192.16-3.82, which is associated with the giant Herbig-Haro flow HH\,396/397, spanning over 10~pc.}{We present new, high spatial and spectral resolution observations of the complex at near-infrared wavelengths ($2.0-2.3$~$\mu$m) using the LUCI near-infrared camera and spectrograph with the Advanced Rayleigh guided Ground layer adaptive Optics System, ARGOS, at the Large Binocular Telescope.}{We discover a string of tightly collimated knots of H$_2$ emission, spanning the full observed field of $\sim4$\arcmin{}, and determine an excitation temperature of $2600\pm500$~K for the brightest knot, which is situated close to the driving source.  We show that the kinematics of the knots are consistent with them being ejected from the central source on timescales of a few times $10^{2-3}$ years.}{The driving source (or sources) of the outflow is obscured at near-infrared wavelengths, possibly due to a thick accretion disk. The distribution of H$_2$ emission in the region, together with high mass-infall rates reported recently, indicate G192 has undergone several large accretion bursts in the recent past.}
\maketitle

\section{Introduction}

Because of the large extinctions and distances typically involved, direct observations of the accretion process of mass onto forming massive stars are difficult to obtain.  However, the accretion and ejection processes in massive young stellar objects (MYSOs) are known to be linked, as shown from observational studies of molecular outflows at radio wavelengths \citep[e.g.][]{Beuther02,Maud15} and collimated jets at infrared wavelengths \citep[e.g.][]{Caratti15}, and further supported by numerical simulations by \citet{Kolligan18}.

In this Letter, we present new high-resolution near-infrared observations of the environment around the massive young stellar object G192.16-3.82 (G192).  Radio observations by \citet{Hughes93} of the ultra compact \ion{H}{II} region associated with this MYSO indicate it harbors a forming early B-star, with a luminosity of $\sim 2400 L_\odot$ scaled to the  distance of $1.52 \pm 0.08$~kpc found by \citet{Shiozaki11}. Early interest in the 1990s was raised because this region exhibits a strong and wide bipolar CO outflow \citep{Shepherd98}. Traces of (shocks in) the bipolar outflow can even be seen in the optical, where H$\alpha$ and [\ion{S}{II}] emission associated with the Herbig-Haro objects HH\,396/397 can be found over a total extent of $>$~5 pc in each direction from the driving source \citep{Devine99}.

 \citet{Imai06} discovered several H$_2$O maser features in the G192 region. Two of the three maser clusters are associated with the infalling/rotating accretion disk of the northernmost YSO, and the third with the highly collimated jet of the southern YSO. The distance between them is about 1200~au.  Using VLA observations with a spatial resolution of 40~mas, \citet{Shepherd01} established the presence of a solar-system-sized circumstellar dust disk around the southern YSO. Moreover, they suggested the presence of a close companion of the southern source, located at a distance of $\sim80$~au. These observations brought G192 to wide attention within the star-formation community and indicated that at least early B-stars may form in a general accretion-disk scenario like their lower-mass siblings, the T~Tauri and Herbig Ae stars. Clear observational support for this is still rare in case of even more massive O-stars.

At near-infrared wavelengths, the intricate morphology of the close circumstellar environment of G192 has proven difficult to discern given the resolution limits of existing observations.
\citet{Indebetouw03} and \citet{Varricatt10} presented near-infrared narrow-band imaging of the region. The latter work identified a few knots of H$_2$ emission away from the central source, however, these seeing-limited observations illustrate the limits of 3-4 meter class telescopes in recovering faint extended emission in such distant star-forming regions. High spatial resolution and signal-to-noise data are needed to study the substructure and dynamics of the region.  In this letter, we present a near-infrared investigation with 0.25\arcsec{} spatial resolution and $R\sim10000$ spectral resolution of the G192 region.

\section{Observations and data reduction}
\label{Sect:ObsReduct}

We observed G192
in the near-infrared with the LUCI imager and spectrograph \citep{Seifert03,Buschkamp12} at the Large Binocular Telescope (LBT) on Mount Graham International Observatory between 2015 and 2017, during several commissioning nights of the Advanced Rayleigh guided Ground layer adaptive Optics System \citep[ARGOS;][]{Rabien19}. These ground-layer-corrected AO observations cover the entire 4\arcmin{}$\times$4\arcmin{} of the LUCI field of view, including G192 and extended outflows in the region.

\subsection{Imaging}

Observations in the $K_s$, H$_2$ and Br$\gamma$ filters were obtained
on December 17th, 2015, with LUCI1.  The observing strategy consisted of small dither offset exposures (to remove bad pixels) shortly followed by a large offset set of exposures to build a sky frame. Each exposure was of 3~sec in order to minimize saturation and persistence effects. The latter we corrected on a pixel-by-pixel basis with persistence and linearity maps \cite[see Appendix~A in][]{Georgiev19}, prior to performing standard dark, flat fielding, sky subtraction and final image registration. Finally, the $K_s$ image is an average combined from 58 exposures resulting in 174~sec of total exposure time. The H$_2$ and Br$\gamma$ filters were combined from 72 and 39 exposures, respectively, each of 6~s, for total on-source exposure times of 432~sec (H$_2$) and 234~sec (Br$\gamma$). The DIMM seeing at $\lambda\!=\!0.55$~$\mu$m was about 1.0\arcsec{}, corresponding to about 0.76\arcsec{} at $\lambda\!=\!2.2$~$\mu$m. The pixel scale of the images is 0.118\arcsec{}/pix, and the FWHM of the PSF after AO correction is 0.31\arcsec{} in both
narrow-band filters and 0.37\arcsec{} in $K_s$.

We obtained an astrometric solutions with
Astrometry.net \citep{Lang10}, using
index files compiled from the Gaia DR2 catalog \citep{GaiaDR2}. The accuracy of the absolute astrometric solution, estimated by measuring the scatter in the positions of the approximately 100 Gaia stars in the
field, is 79~mas in the H$_2$ and Br$\gamma$ filters, and 98~mas in the $K_s$ filter.

\subsection{Spectroscopy}

Long-slit spectroscopic observations with LUCI1\&2 were performed using both 8-m telescopes in binocular mode
on the night of March 11th, 2017 with a spectral resolution of $R\approx 10\,000$ using the 0.25\arcsec{} slit.  We took a total of eight 150~s exposures in sequence, including two off-target positions for sky subtraction, for a total on-source integration time of 15 minutes. The slit was oriented at a position angle of 92.5\degr{}, covering the position of the 1.3-cm continuum source and several H2 knots visible in the narrow-band H2 images.  The wavelength coverage was 2.104-2.256~$\mu$m and 2.151-2.303~$\mu$m for LUCI1 and LUCI2, respectively.

The spectra were reduced in IDL using the Flame data reduction pipeline \citep{Belli18}, including flat fielding, wavelength calibration and field distortion correction using telluric OH lines, and corrections for non-linearity of the detector.  The spectra from LUCI1 and LUCI2 were averaged together to increase the signal-to-noise ratio in the common wavelength range (2.151-2.256~$\mu$m).  No absolute flux calibration was applied.

\section{Results}\label{Sect:Results}

\subsection{Morphology}

\begin{figure}
\centering
\includegraphics[width=\hsize]{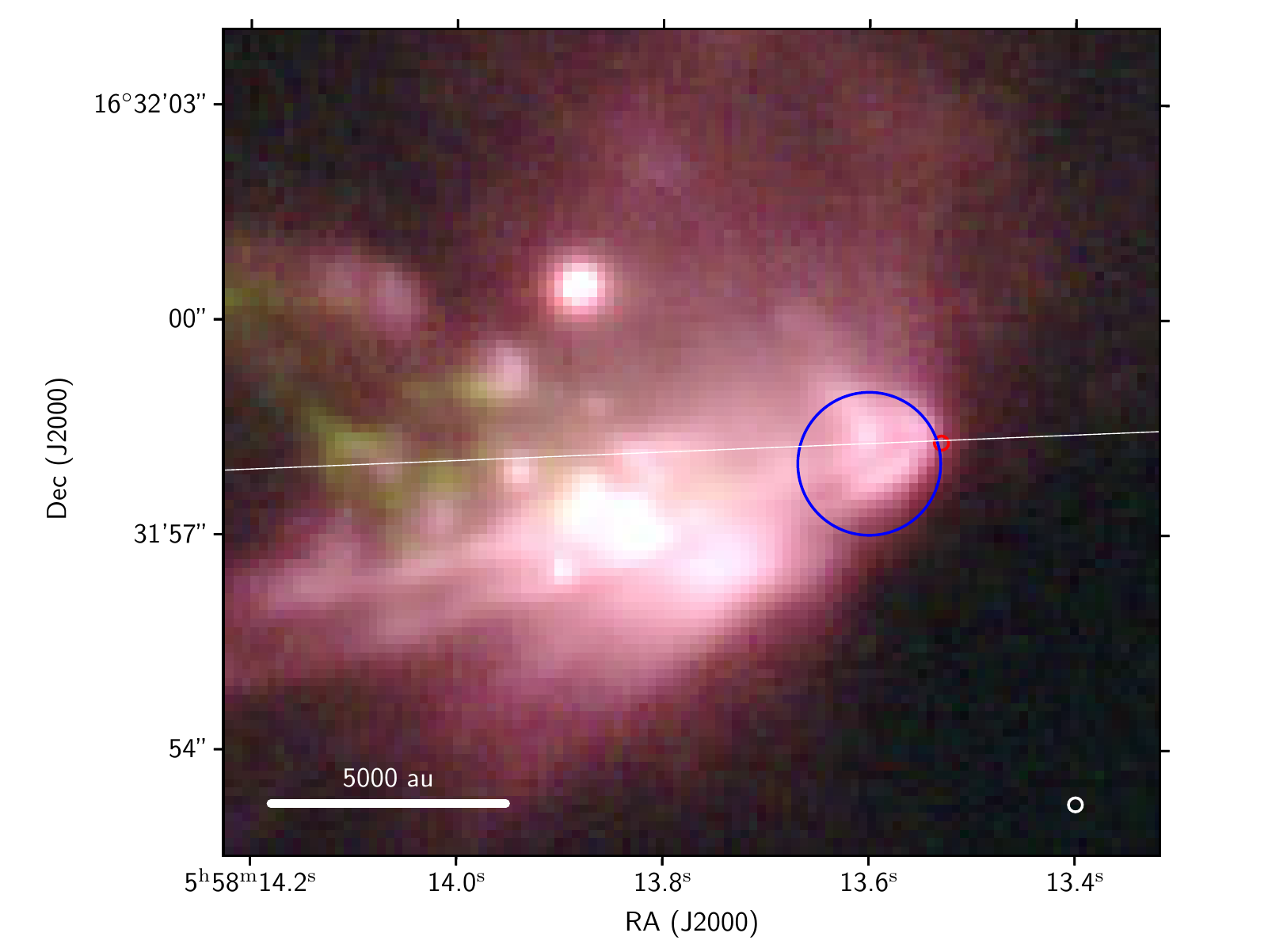}
\caption{Color-composite image of the inner region of G192 showing emission in the $K_s$ (red), H$_2$ (green) and Br$\gamma$ (blue) filters. The blue circle marks the position and error ellipse of the $K^\prime$ source detected by \citet{Indebetouw03}, and the red circle marks the location of the 1.3-cm continuum source \citep{Shepherd04} with an assumed positional uncertainty $\le0.1$\arcsec{}. The white circle in the lower right shows the positional accuracy (98~mas) of our astrometric solution, and the thin white line shows the slit position of the spectroscopic observations.}
\label{fig_Kimage}
\end{figure}

\begin{figure*}
\centering
\includegraphics[width=\hsize]{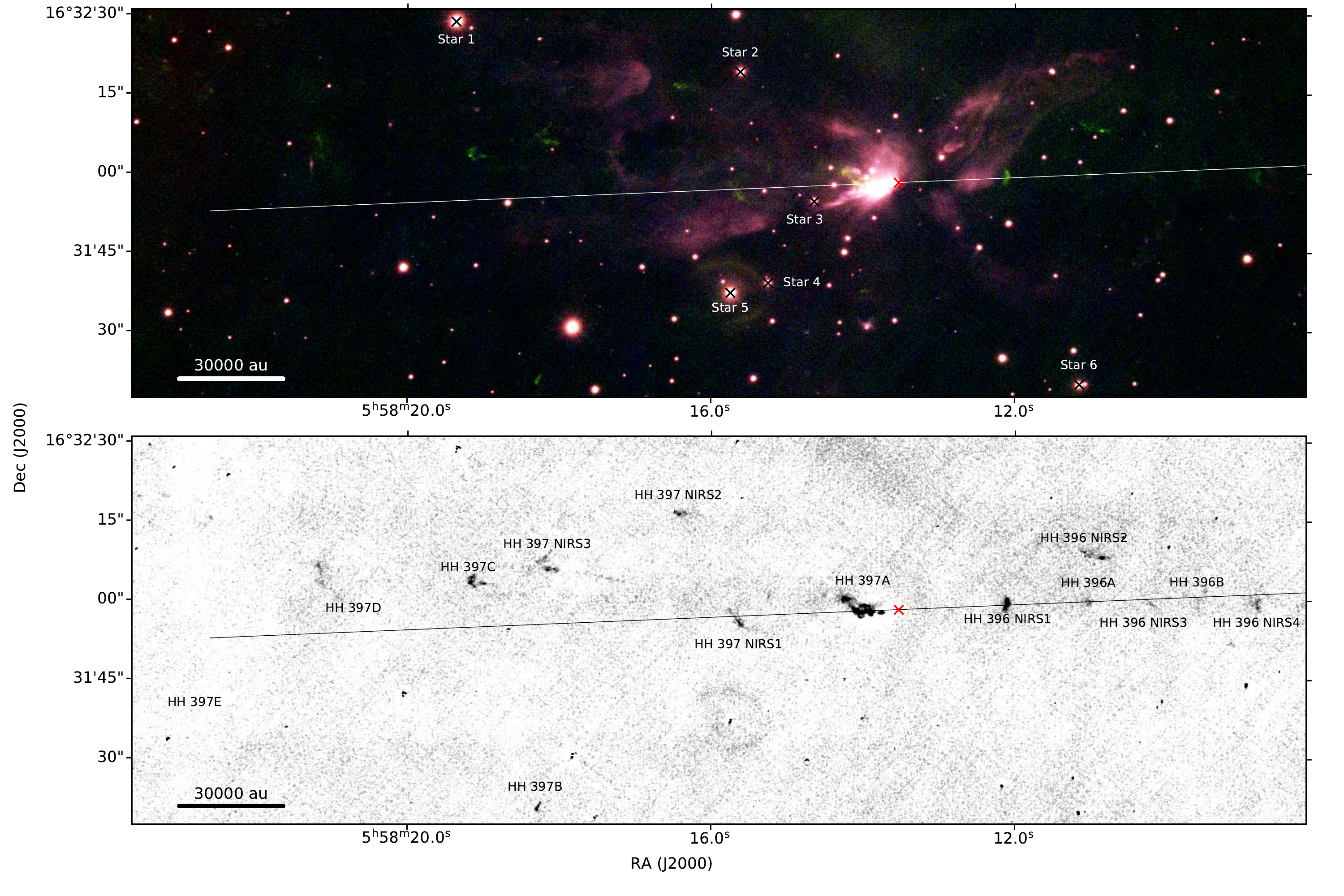}
\caption{\emph{Above}: Color-composite image showing emission in the $K_s$ (red), H$_2$ (green) and Br$\gamma$ (blue) filters. \emph{Below}: Continuum-subtracted H$_2$ image.  The slit position of the spectroscopic observations is overlaid in both panels.  Stars observed by \citet{Jones04} are marked in the upper panel, while individual knots in HH\,396/397 (A, B, etc.) identified by \citet{Devine99} are marked in the lower panel, together with near-infrared knots (NIRS1, 2, etc.) identified in the present work.}
\label{fig_H2-cont}
\end{figure*}

Figure~\ref{fig_Kimage} shows a color-composite image of the $K_s$, H$_2$ and Br$\gamma$ emission of the inner region (12\arcsec{}$\times$14\arcsec{}) around G192, together with the locations and positional uncertainties of the 1.3-cm continuum source of \citet{Shepherd98} and the $K^\prime$ source reported by \citet{Indebetouw03}.  In Fig.~\ref{fig_H2-cont}, we show the large-scale $K_s$, H$_2$ and Br$\gamma$ emission in the region, together with the continuum-subtracted H$_2$ emission. In the continuum-subtracted image the elongated structure of several H$_2$ knots, some of which are associated with optical emission from HH 396/397 (HH\,396A, B; HH\,397A, B, C, D), become visible.  The knots to the east of the 1.3-cm continuum source are associated with the blue-shifted lobe of the outflow, while those to the west are associated with the red lobe (see Sec.~\ref{sec_kinematics}). We have designated knots detected primarily in the near-infrared as NIRS1, NIRS2, etc. These knots are part of the large-scale molecular outflow moving along the east-west direction.  The western knots lie along a relatively straight line, and are spaced at roughly equal intervals of 10-15\arcsec.  To the east, several knots (HH\,397~NIRS3, C, D) show extended arc-like features, resembling bow shocks. The east-west orientation seen in the near infrared matches that of the CO and optical outflows \citep{Shepherd98,Devine99}.  Additionally, a faint ``bubble'' of H$_2$ emission is seen around Star~5 from \citet{Jones04}.

\subsection{Kinematics}
\label{sec_kinematics}

\begin{figure*}
\centering
\includegraphics[width=0.9\hsize]{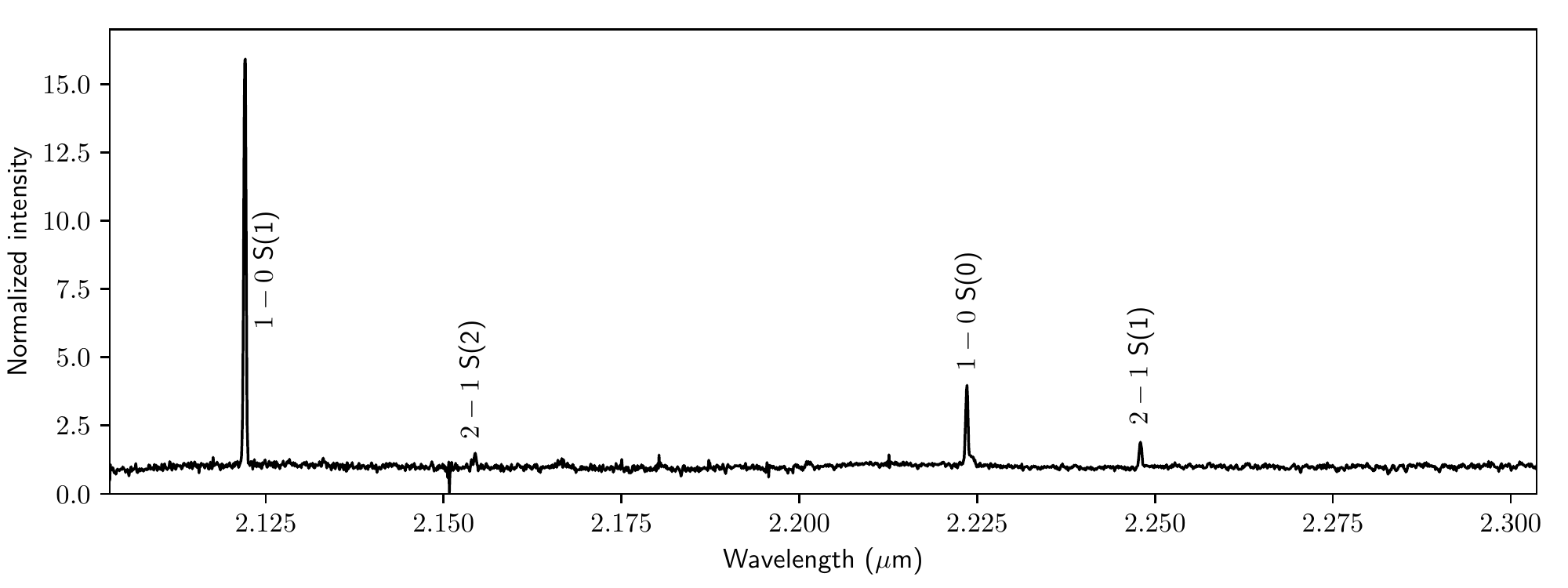}
\caption{Continuum-normalized one-dimensional spectrum of HH\,397A, averaged over 9\arcsec{}, where the detected H$_2$ emission lines are labeled.}
\label{fig_1dspect}
\end{figure*}

\begin{figure*}
\centering
\includegraphics[width=0.9\hsize]{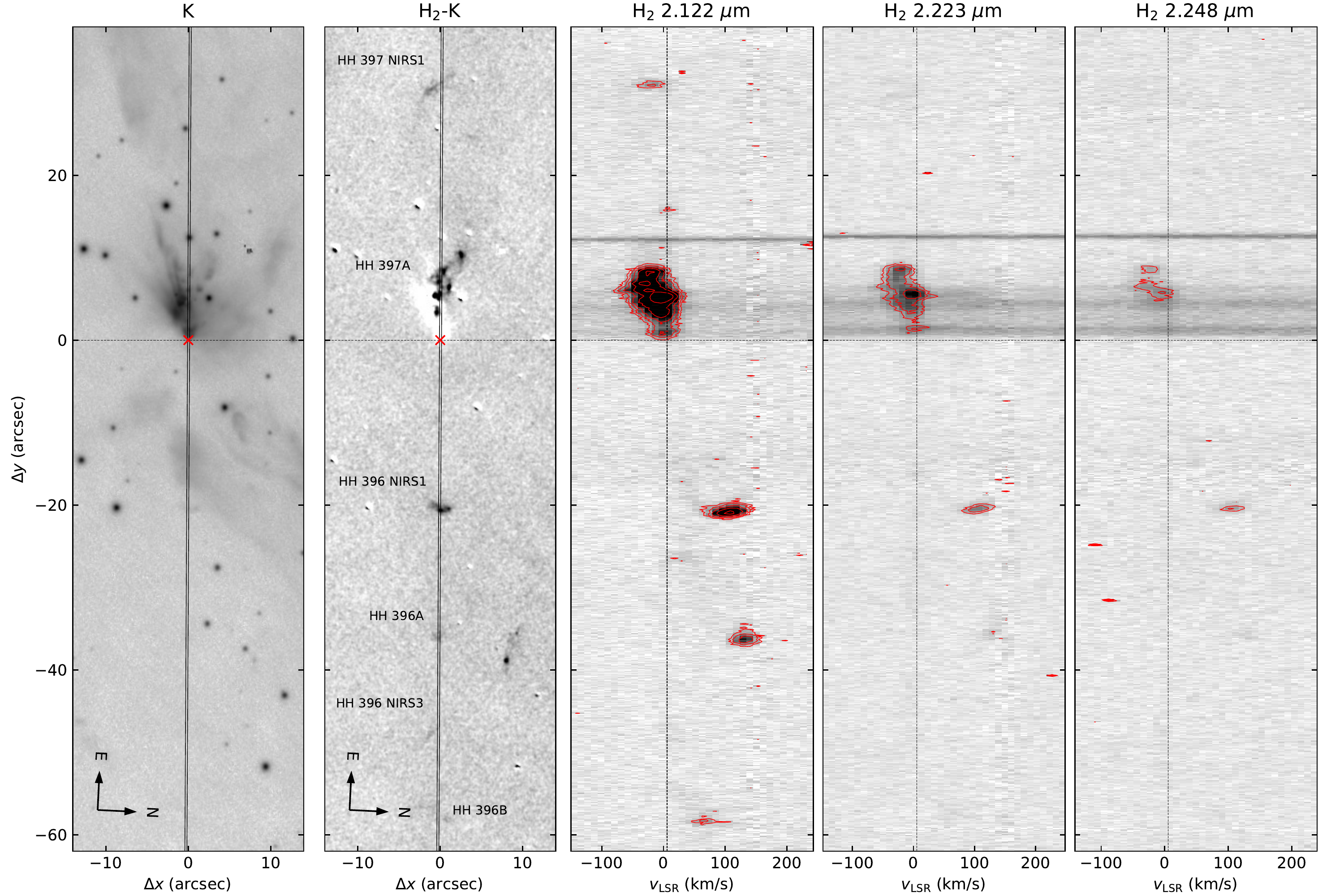}
\caption{{\it Left two panels}: emission in the $K_s$ filter and H$_2$-$K_s$ (continuum subtracted) with the slit position overlaid (solid black lines).  The red cross marks the position of the 1.3-cm continuum emission \citep{Shepherd04}, which is placed at the origin of the coordinates $\delta x$ (across the slit) and $\delta y$ (along the slit). {\it Right three panels}: PV diagrams of the three lines of H$_2$ emission along the slit. Contours have been drawn for the continuum-subtracted H$_2$ emission at 4.5, 8.2, 15, 27 and 50 times the background rms level. The horizontal dotted lines mark the location of the radio continuum source along the slit in all five panels, while the vertical dotted lines in the right three panels show the velocity of the peak of NH$_3$ emission, 5.7~km~s$^{-1}$ (LSR), reported by \citet{Shepherd04}.}
\label{fig_pv}
\end{figure*}

To obtain two-dimensional spectra, the spectrograph slits of both telescopes were oriented along the western episodic outflows, as shown in Fig.~\ref{fig_H2-cont}. The brightest emission is in the H$_2$ lines $1-0$~S(1) (2.12~$\mu$m), $1-0$~S(0) (2.22~$\mu$m), and $2-1$~S(1) (2.24~$\mu$m), with weak emission also detected from HH\,397A in the $2-1$~S(2) (2.154~$\mu$m) line (Fig.~\ref{fig_1dspect}).  Notably, no Br$\gamma$ emission was detected in HH\,397A, which is at odds with the narrow-band observations presented by \citet{Indebetouw03}.  HH\,397A has a bright, red continuum contribution (see Fig.~\ref{fig_pv}), which is not seen in any of the other knots. The continuum has an irregular spatial structure along the slit, with two bright emission peaks about 3\arcsec{} (4500~au) apart.

\begin{figure}
\centering
\includegraphics[width=\hsize]{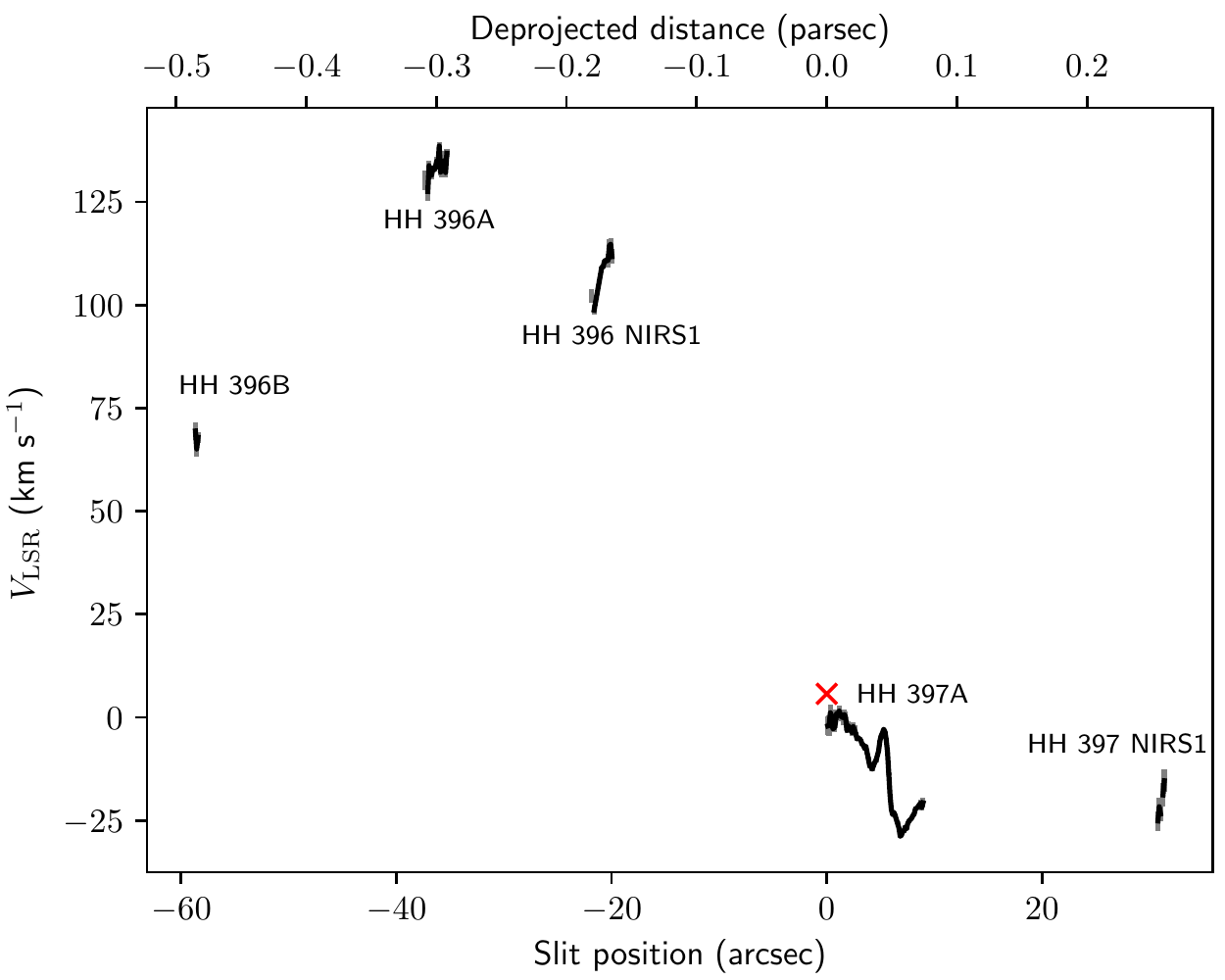}
\caption{Velocity of H$_2$ emission along the slit obtained from fitting the line profiles.  The red cross marks the location of the 1.3-cm continuum source and the peak velocity of NH$_3$ emission \citep{Shepherd04}.}
\label{fig_vel}
\end{figure}

The three right panels in Fig.\,\ref{fig_pv} show PV diagrams of the H$_2$ emission along the slit. The knots appear to be episodic in nature, and have an asymmetrical velocity structure.  Each western knot is red shifted relative to the peak of NH$_3$ emission reported by \citet{Shepherd04}, while the eastern knots are blue shifted.  We fit the line profiles with a Gaussian at each point along the slit to determine the radial velocity (Fig.\,\ref{fig_vel}), and show the mean radial velocity of each knot in Table\,\ref{tab_knots}.  Finally, using an inclination angle of the outflow of 63\degr{} from the line of sight \citep{Shepherd98}, we estimate the corresponding proper motion, along with the back-projected launch time from the assumed starting location, the position of the 1.3-cm source, of each knot observed in the slit (see Table\,\ref{tab_knots}).

\begin{table*}
\caption{Parameters of H$_2$ emission knots in G192}
\label{tab_knots}
\centering
\begin{tabular}{l l l c c c}
\hline\hline

Knot & RA & Dec &$\langle v_\mathrm{LSR} \rangle$ & Estimated proper motion & Launch time \\
& (J2000) & (J2000) & (km s$^{-1}$) & (mas yr$^{-1}$) &  (yrs ago) \\

\hline                                   
HH 396 NIRS4 & 5$^\mathrm{h}$58$^\mathrm{m}$8.8$^\mathrm{s}$ & +16\degr{}31\arcmin{}59.3\arcsec{} & & & \\
HH 396B & 5$^\mathrm{h}$58$^\mathrm{m}$9.6$^\mathrm{s}$ & +16\degr{}31\arcmin{}59.8\arcsec{} & $67 \pm 3$ & $\sim16$ & $\sim4000$ \\
HH 396 NIRS3 & 5$^\mathrm{h}$58$^\mathrm{m}$10.2$^\mathrm{s}$ & +16\degr{}31\arcmin{}59.2\arcsec{} & & & \\
HH 396A & 5$^\mathrm{h}$58$^\mathrm{m}$11.0$^\mathrm{s}$ & +16\degr{}31\arcmin{}59.7\arcsec{} & $133 \pm 6$ & $\sim32$ & $\sim1300$ \\
HH 396 NIRS2 & 5$^\mathrm{h}$58$^\mathrm{m}$11.1$^\mathrm{s}$ & +16\degr{}32\arcmin{}7.0\arcsec{} & & & \\
HH 396 NIRS1 & 5$^\mathrm{h}$58$^\mathrm{m}$12.1$^\mathrm{s}$ & +16\degr{}31\arcmin{}59.3\arcsec{} & $107 \pm 23$ & $\sim25$ & $\sim900$ \\
HH 397A & 5$^\mathrm{h}$58$^\mathrm{m}$13.8$^\mathrm{s}$ & +16\degr{}31\arcmin{}57.1\arcsec{} & $ -24 - 0$ & $\sim0 - 6$ & $\lesssim2100$ \\
HH 397 NIRS1 & 5$^\mathrm{h}$58$^\mathrm{m}$15.6$^\mathrm{s}$ & +16\degr{}31\arcmin{}55.7\arcsec{} & $-20 \pm 12$ & $\sim5$ & $\sim7000$ \\
HH 397 NIRS2 & 5$^\mathrm{h}$58$^\mathrm{m}$16.4$^\mathrm{s}$ & +16\degr{}32\arcmin{}16.3\arcsec{} & & & \\
HH 397 NIRS3 & 5$^\mathrm{h}$58$^\mathrm{m}$18.2$^\mathrm{s}$ & +16\degr{}32\arcmin{}5.7\arcsec{} & & & \\
HH 397B & 5$^\mathrm{h}$58$^\mathrm{m}$18.3$^\mathrm{s}$ & +16\degr{}31\arcmin{}21.0\arcsec{} & & & \\
HH 397C & 5$^\mathrm{h}$58$^\mathrm{m}$19.2$^\mathrm{s}$ & +16\degr{}32\arcmin{}3.1\arcsec{} & & & \\
HH 397D & 5$^\mathrm{h}$58$^\mathrm{m}$20.7$^\mathrm{s}$ & +16\degr{}31\arcmin{}55.9\arcsec{} & & & \\

\hline                                   

\end{tabular}
\tablefoot{Mean radial velocities and corresponding proper motion estimates were determined only for knots lying along the slit.}
\end{table*}

\subsection{Excitation temperature of HH\,397A}

\begin{figure}
\centering
\includegraphics[width=\hsize]{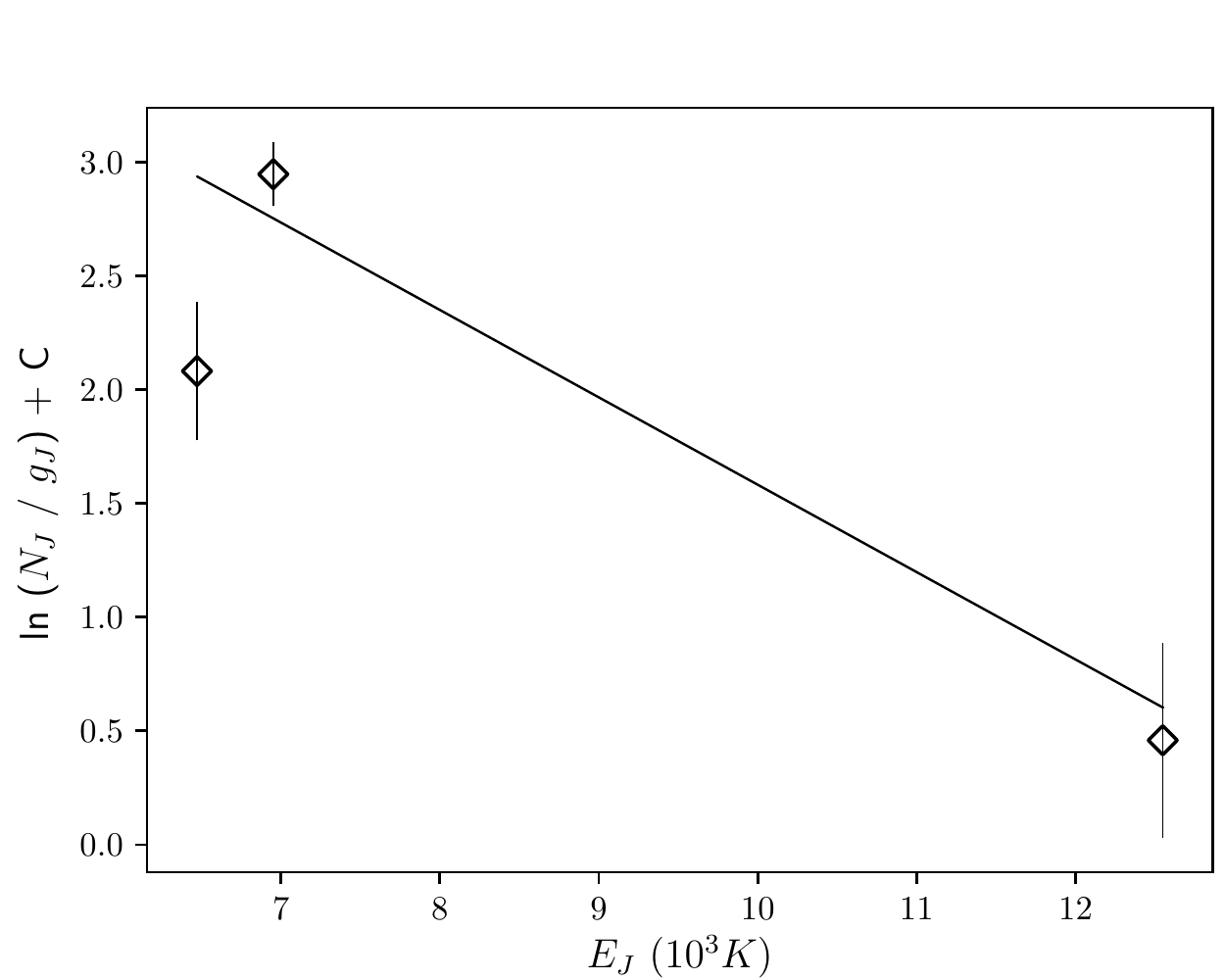}
\caption{Boltzmann diagram for the $2-1$~S(1) (2.24~$\mu$m), $1-0$~S(0) (2.22~$\mu$m) and $1-0$~S(1) (2.12~$\mu$m) H$_2$ lines (left to right) in the HH\,397A region.  The solid line shows the best fit to the level populations, with an excitation temperature of $2600$~K.}
\label{fig_boltz_temp}
\end{figure}

Under the assumption that conditions are close to local thermodynamic equilibrium, we used the Boltzmann distribution to estimate the excitation temperature of HH\,397A. In Fig.~\ref{fig_boltz_temp} we plot the logarithm of the H$_2$ column density $N_{\nu}$ (plus an arbitrary constant) against the corresponding energy level $E_{\nu}$ for each line. The column density may be expressed as $N_J = \cfrac{2 F_J \lambda_J}{A_J \hslash c}$, where $F_J, \lambda_J$ and $A_J$ are the flux, line wavelength and Einstein coefficient of each transition, respectively. Since our observations are not absolute flux calibrated, the value of the column density can't be calculated directly; however, the Boltzmann distribution requires only the ratios of fluxes to estimate the excitation temperature (i.e. $\ln{\cfrac{N_J}{g_J}} \propto \cfrac{E_J}{T}$, where $g_J, E_J$ and $T$ are statistical weight of the transition, energy of upper level and excitation temperature, respectively). We estimate the relative flux calibration uncertainty to be on the order of 10\%, and ignore the effects of interstellar absorption on the line fluxes, which are small due to the closeness in wavelength and energy levels of the transitions. Under these assumptions, we find an average excitation temperature in the HH\,397A region of $2600\pm500$~K.

\section{Discussion}
\label{Sect:Discussion}

\citet{Indebetouw03} identified a very red 2-$\mu$m source near the position of the millimeter source of \citet{Shepherd01}.  Given their seeing-limited spatial resolution and positional accuracy of $\sim1$\arcsec{}, they concluded that the 2-$\mu$m source and millimeter sources are coincident, with the emission at 2~$\mu$m arising from a combination of photospheric emission and scattered light from the driving source(s) in G192.  Using arguments based on line-of-sight extinction to the central source, they speculate that the circumstellar disk of G192 may have a cleared inner hole (roughly $10-15$~au), and that active accretion in the source has ended or temporarily stopped.

Our AO-corrected imaging and much higher positional accuracy show that the 2~$\mu$m and millimeter emission in G192 are distinct, with no significant near-infrared emission at the position of the millimeter source.  Indeed, as \citet{Indebetouw03} show, given the estimates for the disk mass \citep[$3\!-\!15$\,M$_\odot$;][]{Shepherd01} and inclination angle of the outflow \citep[$\sim$63\degr;][]{Shepherd98}, there should be no detectable near-infrared emission if we are looking through the mid-plane of the disk.  At the same time, the relatively bright continuum component in HH\,397A, which is absent in all the other near-infrared extended sources in the region, indicates that this region is indeed illuminated by one or more YSOs in the vicinity.  The derived excitation temperature of $\sim\!2600$\,K~in HH\,397A is consistent with shock excitation of the gas, indicating that this region is not merely a quiescent reflection nebula.  Taken together with the age estimates for the other knots in the region, this implies that accretion bursts in G192 have been taking place quasi-regularly over at least the past few $10^2\!-\!10^3$ years.

The velocity distribution of the red-shifted knots shows that the HH\,396 NIRS1 and NIRS2 knots have higher velocities than the HH\,396 NIRS3 knot, which may be an indication of interactions with interstellar surrounding material.  This same scenario may be occurring in the case of HH\,397 NIRS1 in the eastern (blue-shifted) blue part of the outflow, if the outflow material was ejected into a region with much higher local density, and hence almost instantly decelerated to the low velocities ($\sim20$~km~s$^{-1}$) observed.  On the other hand, another explanation of such a velocity distribution may be related to deceleration of accretion, where the slowest (and closest to the central source) knot might have been ejected when the accretion rate was lower.

An explanation of apparent episodic nature of the outbursts may be related to the fragmentary structure of the accretion disk. \citet{Meyer17} performed 3D radiation hydrodynamics simulations, showing the gravitational collapse of a 100~M$_\odot$ proto-stellar core, tracing the evolution of the accretion disk over  30~kyr. The accreting material around the protostellar core forms into spiral arms, and density inhomogeneities cause part of the material in the arms to fragment into clumps. These clumps continue to fall onto the star, causing periodic outbursts of matter. The high infall rates recently derived for G192 by \citet{Tang19} of $>10^{-3}$ M$_\odot$/yr support this picture. These SMA molecular line measurements trace the infall from the surrounding envelope to the central (potentially disk-like) part of G192, and such high infall rates can certainly induce instabilities within the disk, which could lead to subsequent fragmentation.

It is interesting to note that the blue-shifted velocities in the strongest H$_2$ emission structure, HH397A, just raise up to around $-30$~km/s. In comparison, the SMA CO mapping by \citet{Liu13} revealed a high-velocity CO component at this location with velocities of over $-70 ... -90$~km/s. This can be reconciled when assuming that the inner parts of the molecular outflow have a strong component with a large opening angle of $\sim 90^\circ$ \citep{Shepherd98, Shepherd99}. Hence, the CO observations include a component with smaller inclinations to the observer, while the H$_2$ flow mainly proceeds along the main jet axis, $\sim 63^\circ$ inclined to the line-of-sight.

Finally, we note that for the proper motions of the knots estimated in Table~\ref{tab_knots}, observations with comparable spatial resolution ($\sim300$~mas) and astrometric accuracy ($\sim100$~mas) should reveal positional changes of the outflow knots on the order of 100~mas after only five years.

\section{Summary and conclusion}

In this work, we presented new AO-corrected photometric and spectroscopic observations of a 4\arcmin{} field around the massive young stellar object G192.16-3.82 at near-infrared wavelengths ($2.0-2.3$~$\mu$m).

Using continuum-subtracted narrow-band images, we revealed several knots of emission in the H$_2$ line at 2.12~$\mu$m lying approximately along the east-west axis, consistent with the morphology of the Herbig-Haro flow seen at optical wavelengths.  Five of the six knots detected to the west of the central MYSO lie along a straight line and are spaced roughly evenly at intervals of $10-15$\arcsec{}.  The back-projection launch times of the western knots, which were presumably associated with major accretion events, were estimated from the observed radial velocities and assumed inclination of 63\degr{} and span up to $\sim10^3-10^4$~yrs in the past.  The H$_2$ emission to the east of the MYSO, on the other hand, is more irregular. The morphologies of several of the knots resemble bow shocks, with the brightest line and continuum emission arising in HH~397A near the central source.  The excitation temperature derived from the H$_2$ line fluxes in this region is consistent with shock excitation.

The high spatial resolution ($0.3-0.4$\arcsec) and precision of the astrometric solution ($0.08-0.1$\arcsec) of the observations allowed us to determine that the driving source (or sources) of the large-scale outflow HH~396/397 remains completely obscured at near-infrared wavelengths, which suggests that the line of sight to the central source lies through the midplane of the disk.  Taken together with the high infall rates, the observed asymmetries and episodic nature of the G192 outflow at near-infrared wavelengths support the picture that the central MYSO is still accreting material on scales of hundreds to thousands years, and that the accretion disk may have undergone several fragmentation incidents in the recent past.

\begin{acknowledgements}
We are grateful for the support of Martin Kulas, Jose Borelli, Diethard Peter, Julian Ziegeleder, Tommaso Mazzoni for technical support with ARGOS and gathering data in the early stages of instrument commissioning. We thank Debra Shepherd for discussions and for providing her OVRO interferometric CO data in electronic form.
The work of P. Boley and N. Dmitrienko was supported by grant 18-72-10132 of the Russian Science Foundation. 
This research made use of Astropy\footnote{http://www.astropy.org}, a community-developed core Python package for Astronomy \citep{astropy13, astropy18}, as well as Matplotlib \citep{Hunter07}.
The LBT is an international collaboration among institutions in the United States, Italy and Germany. LBT Corporation partners are: The University of Arizona on behalf of the Arizona university system; Instituto Nazionale di Astrofisica, Italy; LBT Beteiligungsgesellschaft, Germany, representing the Max-Planck Society, the Astrophysical Institute Potsdam, and Heidelberg University; The Ohio State University, and The Research Corporation, on behalf of The University of Notre Dame, University of Minnesota and University of Virginia.
\end{acknowledgements}

\bibliography{ms}

\end{document}